\documentclass[useAMS,usenatbib,referee]{mn2e}{{
\usepackage{graphicx}
\usepackage{natbib}
\include{epsf}
\title[]{The effect of grain size distribution on H$_2$ 
formation rate in the interstellar medium
}

\author[A. Lipshtat and, O. Biham] {Azi Lipshtat$^{1}$ and Ofer 
Biham$^{1}$  \\  
$^{1}$Racah Institute of Physics, The Hebrew University, Jerusalem 
91904, Israel \\
}

\begin{document}
\maketitle

\begin{abstract}

The formation of molecular hydrogen in the interstellar medium takes place
on the surfaces of dust grains. 
Hydrogen molecules play a role in gas-phase reactions that produce other
molecules, some of which serve as coolants
during gravitational collapse and star formation. 
Thus, the evaluation of the production rate of hydrogen molecules
and its dependence on the physical conditions 
in the cloud are of great importance.
Interstellar dust grains exhibit a broad size distribution
in which the small grains capture most of the surface area.
Recent studies have shown that the production efficiency strongly depends on
the grain composition and temperature as well as on its size. 
In this paper we present a formula which 
provides the total production rate of H$_2$ per unit volume in the cloud,
taking into account the
grain composition and temperature as well as the grain
size distribution.
The formula agrees very well with the master equation results.
It shows that for a physically relevant range of grain temperatures, the
production rate of H$_2$ is significantly enhanced 
due to their broad size distribution.

\end{abstract}


\section{Introduction}
        \label{intro.}
 
The chemistry of interstellar clouds consists of reactions taking place in
the gas phase as well as on the surfaces of dust grains
\citep{Hartquist1995}.
In particular, 
the formation of H$_2$ takes place on grain surfaces
while all competing gas phase processes are orders of 
magnitude less efficient
\citep{Gould1963}.
Hydrogen molecules are necessary  for the initiation of gas-phase reaction
networks that give rise to the chemical complexity observed in interstellar  
clouds. Molecules serve as coolants during
gravitational collapse, enabling the formation of stars.

In diffuse clouds the H$_2$ formation process takes place on
bare grains that consist of amorphous silicates and carbon.
In dense molecular clouds the grains are coated by ice mantles.
Upon collision with a grain, an
H atom has a probability $\xi$ to become adsorbed on the surface.
The adsorbed H atom (adatom)
resides on the surface for an average time 
$t_{\rm H}$ (residence time)
before it desorbs.
In the Langmuir-Hinshelwood mechanism,
the adsorbed H atoms diffuse on the surface of the grain
either by thermal hopping or by tunneling. 
When two H atoms encounter each other on the surface, 
an H$_2$ molecule may form 
\citep{Williams1968,Hollenbach1970,Hollenbach1971a,Hollenbach1971b,Smoluchowski1981,Smoluchowski1983,Aronowitz1985,Duley1986,Pirronello1988,Sandford1993}.
The formation rate 
$R_{\rm total}$ (cm$^{-3}$s$^{-1}$) 
of molecular hydrogen 
on dust-grain surfaces
is commonly evaluated using the
Hollenbach-Werner-Salpeter formula 

\begin{equation}
R_{\rm total} = 
{1\over 2} n_{\rm H}\langle v_{\rm H}
\rangle n_{\rm grain} \langle \sigma_g\rangle\gamma,
\label{eq:HSlong}
\end{equation}

\noindent
where $n_{\rm H}$  (cm$^{-3}$) 
is the number density of hydrogen atoms in the gas,
$\langle v_{\rm H}\rangle$ (cm s$^{-1}$) 
is the average velocity of hydrogen atoms,
$n_{\rm grain}$ (cm$^{-3}$) 
is the number density of dust grains
and
$\langle \sigma_g\rangle$ (cm$^2$)
is the average grain cross section
\citep{Hollenbach1971b}. 
Assuming spherical grains, 
$\langle \sigma_g\rangle = \pi \langle r^2 \rangle$ 
where $r$ (cm) is the grain radius.
The parameter  
$0 \le \gamma \le 1$ 
is the recombination efficiency,
namely, the fraction of atoms hitting the grain which 
come out in molecular form.

The total grain mass
amounts to about $1\%$ of the 
hydrogen mass in the cloud.
The number density of dust grains is about 
$10^{-13}-10^{-12}$ smaller than the total number 
density of hydrogen atoms. 
In previous studies, a
typical grain radius of $0.17 \mu$m
was used, assuming a mass density of 
$\rho_g = 2$ (gram cm$^{-3}$)
\citep{Hollenbach1971b}.
Under these assumptions, for
gas temperature of 100 K
and $\gamma = 0.3$,
Eq. ~(\ref{eq:HSlong})  
can be simplified to 

\begin{equation}
R_{\rm total} = R \cdot n_{\rm H} n,
\label{eq:astronomers}
\end{equation}

\noindent
where
$R \simeq 10^{-17}$ (cm$^3$ s$^{-1}$)
is the rate coefficient and
$n=n_{\rm H} + 2 n_{\rm H_2}$ 
(cm$^{-3}$) 
is the total density of hydrogen atoms
in both atomic and molecular ($n_{\rm H_2}$) form
\citep{Hollenbach1971b}. 
Eq.~(\ref{eq:astronomers})
is commonly used 
for the evaluation of the hydrogen recombination rate 
in models of interstellar chemistry.
In this paper we present a direct calculation of
$R_{\rm total}$, without using rate coefficients.

The grain size distribution 
$n_g(r)$ (cm$^{-4}$)
in interstellar clouds
is broad
with many small grains and few large grains.
The surface area 
is dominated by the small grains.
This distribution can be approximated by a power law of the form 
\citep{Mathis1977,Weingartner2001} 

\begin{equation}
n_g(r) = c r^{-\alpha} 
\label{eq:distribution} 
\end{equation}

\noindent
where 
$n_g(r) dr$ (cm$^{-3}$)
is the density of grains with radii in the range $(r,r+dr)$
in the cloud. 
This distribution is bounded between the upper cutoff
$r_{\rm max}= 0.25 \mu {\rm m}$ 
and the lower cutoff
$r_{\rm min} = 5 {\rm nm}$. 
The exponent is  
$\alpha \simeq 3.5$ 
\citep{Draine1984}.
In general, 
the normalization constant is given by 

\begin{equation}
c={{3M_g(4-\alpha)}
\over{4\pi\rho_g(r_{\rm max}^{4-\alpha}-r_{\rm min}^{4-\alpha})}},  
\label{eq:normalization_c} 
\end{equation}

\noindent
where
$M_g$ 
(gram cm$^{-3}$)
is the mass density of grain material in the cloud.
The total number density, 
$n_{\rm grain}$ 
(cm$^{-3}$),
of grains is given by

\begin{equation}
n_{\rm grain} = 
\frac{c(r_{\rm min}^{1-\alpha}-r_{\rm max}^{1-\alpha})}{\alpha -1}.
\label{eq:Ng_c} 
\end{equation}

Recent experiments have shown that the mobility of H
atoms on dust analogues under interstellar conditions
is dominated by thermal hopping rather than tunneling
\citep{Pirronello1997a,Pirronello1997b,Pirronello1999,Katz1999,Roser2002}.
As a result, the production efficiency of H$_2$
molecules is highly dependent on the grain temperature.
The efficiency is very high within a narrow 
window along the temperature axis, with exponential decay
on both sides.
Theoretical studies have shown that 
for sufficiently small grains, 
the production efficiency is strongly dependent on the grain size,
namely it decreases as the grain size is reduced
\citep{Biham2001,Green2001}.
The Hollenbach-Werner-Salpeter formula, in which all the surface
processes are captured in the parameter $\gamma$ 
cannot account for the temperature and grain-size dependence
of the production efficiency.

In this paper we present a formula for the 
production rate of H$_2$ in interstellar clouds,
which takes
into account explicitly the dependence of the recombination
efficiency on the grain temperature and size, 
integrating the size dependence over the entire distribution
of grain sizes. 
This formula is in very good agreement with 
the results obtained from the master equation.
It applies for a broad range of steady-state and 
quasi steady-state conditions and enables the evaluation
of the production rate of H$_2$ in a closed form without
the need for any simulation or numerical integration.  

The paper is organized as follows. 
In Sec. 2 we present the different approaches to the calculation of
reaction rates on surfaces.
The dependence of these rates on the grain size and temperature is
considered in Sec. 3.
In Sec. 4 we introduce a formula that accounts for the formation rate
of molecular hydrogen per unit volume of the clouds, 
integrated over the entire distribution of grain sizes. 
The results are 
discussed in Sec. 5 and
summarized in Sec. 6.

\section{Recombination rates on single grains}

Calculations of reaction rates of
H$_2$ formation and other 
chemical reactions on dust grains 
are typically done using rate
equation models
\citep{Pickles1977,Hendecourt1985,Brown1990b,Hasegawa1992,Caselli1993,Willacy1993}. 
These models consist of coupled ordinary differential equations that
provide the time derivatives of the populations of the species involved.
The reaction rates are obtained by numerical integration of these equations.
The production of molecular hydrogen on a single grain 
of radius $r$
is described by

\begin{equation}
{dN \over dt} = F - W N - 2 A N^2,
\label{eq:rate}
\end{equation}

\noindent
where $N$ is the number of H atoms on the grain.  
The first term on the right hand side  
is the flux 
$F$ (sec$^{-1}$) 
of hydrogen atoms
onto the grain surface.
The second term describes the thermal desorption of H atoms from the surface
where

\begin{equation}
W = \nu \exp(-{E_1/k_{\rm B}T})
\label{eq:W}
\end{equation}

\noindent
is the desorption rate,
$\nu$ is the attempt rate 
(standardly taken as $10^{12}\, {\rm sec}^{-1}$),
$E_1$ is the activation energy barrier for desorption and 
$T$ (K) is the grain temperature.
The third term in 
Eq.~(\ref{eq:rate}) 
accounts for the depletion in the number of adsorbed atoms 
due to formation of molecules. 
The parameter $A = a/S$ is the rate at which H atoms scan the entire
surface of the grain, where 

\begin{equation}
a = \nu \exp(-{E_0/k_{\rm B}T})
\label{eq:A}
\end{equation}

\noindent
is the hopping rate between adjacent adsorption sites on the grain,
$E_0$ is the energy barrier for hopping, 
$S=4\pi r^2 s$ 
is the number of adsorption sites on the grain surface and 
$s$ (cm$^{-2}$)
is their density.
The formation rate, 
$R_{\rm grain}$ 
of H$_2$ on a single grain is given by
$R_{\rm grain} = A N^2$.
Under steady state conditions one can obtain an exact solution
for $R_{\rm grain}$
in terms of $F$, $W$ and $A$
\citep{Biham1998}.

Following the experimental results reported by 
\cite{Katz1999}
for the amorphous carbon sample,
we use the parameters:
$s=5\cdot 10^{13}$ (cm$^{-2}$), 
$E_0=44.0$ (meV) 
and 
$E_1=56.7$ (meV).
It is found 
\citep{Biham2002}
that
the process of molecular hydrogen formation
is efficient only 
within a narrow window of grain temperatures,
$T_0 < T < T_1$,
where

\begin{equation} 
T_0 = \frac{E_0}{k_{\rm B} \ln(\nu S/F)}
\end{equation}

\noindent
and

\begin{equation} 
T_1 = \frac{2 E_1 - E_0}{k_{\rm B} \ln(\nu S/F)}.
\end{equation}

\noindent
At temperatures below $T_0$ the
mobility of H atoms on the surface is very low,
sharply reducing the production rate.
At temperatures above $T_1$ most atoms quickly desorb
before they encounter each other and form molecules.

Rate equations are an ideal tool for the simulation of 
surface reactions, due to their simplicity and high computational efficiency.
In particular, they account correctly for the temperature dependence
of the reaction rates.
However, in the limit of small grains under low flux they become unsuitable.
This is because they ignore the fluctuations as
well as the discrete nature of the populations of atoms on the grain
\citep{Charnley1997,Caselli1998,Shalabiea1998,Stantcheva2001}. 
For example, as the number of H atoms on a grain fluctuates 
in the range of 0, 1 or 2,
the H$_2$ formation rate cannot be obtained from the average 
number alone.
This can be easily understood, 
since the recombination process requires at least two
H atoms simultaneously on the surface.

Recently, a master equation approach was proposed, that
takes into account both the discrete nature of the
population of H atoms as well as the fluctuations, and  
is thus suitable for the simulation of H$_2$ formation on
interstellar dust grains
\citep{Biham2001,Green2001}.
Its dynamical variables are the
probabilities $P(N)$ 
that there are 
$N$ 
atoms
on the grain.
The time derivatives  
$\dot{P}(N)$, 
$N=0, 1, 2, \dots$,
are expressed 
in terms of the adsorption, reaction and desorption terms
according to 

\begin{eqnarray}
\dot P(N ) &=& 
  F  \left[ P (N -1) - P (N ) \right] \nonumber \\ 
&+& W  \left[ (N +1) P (N +1) - N  P (N ) \right]  \nonumber \\
&+& A  \left[ (N +2)(N +1) P (N +2) -  N (N -1) P (N ) \right] 
\label{eq:Pmaster}  
\end{eqnarray}

\noindent
where 
$N=0, 1, 2, \dots$.
Note that the equations for 
$\dot P(0)$ 
and
$\dot P(1)$ 
do not include all the terms, because at least one H 
atom is required for desorption to occur and at least two
for recombination.
Direct integration of the master equation provides the time evolution 
of the probabilities 
$P(N)$. 
The recombination rate per grain

\begin{equation}
R_{grain}=A\left(\langle N^2 \rangle - \langle N \rangle\right)
\label{eq:Rgrain}
\end{equation}

\noindent
can then be calculated,
where 
$\langle N^k \rangle$, $k=1,2,\dots$, is the $k$'th moment of the
distribution $P(N)$.
Under steady state conditions the recombination rate 
is given by
\citep{Green2001,Biham2002}

\begin{equation}
R_{\rm grain} = 
{F \over {2}} 
{ {I_{W/A+1} \left(2\sqrt{2 F/A}\right)} 
\over 
{I_{W/A-1} \left(2\sqrt{2 F/A}\right)} },  
\label{eq:Rexact}
\end{equation}

\noindent
where $I_x(y)$ is the modified Bessel function.

Using suitable summations over the master equation, one obtains
the moment equations, namely a set of
coupled differential equations for the time derivatives of the moments 
$\langle N^k\rangle$, $k=1,2,\dots$, 
of the distribution  
$P(N)$  
\citep{Lipshtat2003}. 
The recombination rate on grains of a given size
depends only on
the first two moments, 
$\langle N\rangle$ 
and    
$\langle N^2\rangle$. 
The time derivatives of these moments are given by

\begin{eqnarray}
{d \langle N \rangle \over {dt} } &=& F
+( - W + 2A)  \langle N \rangle 
- 2 A \langle N^2 \rangle  \nonumber \\
{d \langle N^2 \rangle \over {dt} } &=& F
+( 2 F+ W-4A ) \langle N \rangle 
+ (8 A - 2 W) \langle N^2 \rangle
- 4 A \langle N^3 \rangle. 
\label{eq:moment12}
\end{eqnarray}

\noindent
Using suitable cutoffs in the master equation one can replace
the third moment
$\langle N^3 \rangle$
by
$\langle N^{3} \rangle = 3 \langle N^{2} \rangle - 2 \langle N \rangle$,
and obtain a closed set of two coupled differential equations for the
first two moments.
Under steady state conditions Eq.
(\ref{eq:moment12})
takes the form of two coupled linear algebraic equations for
$\langle N \rangle$ 
and
$\langle N^{2} \rangle$.
The solution of these equations is given by
$\langle  N \rangle = { F(A + W) / ({2A F + W A + W^2}) }$ 
and
$\langle  N^2 \rangle = { F(F+A+W) / ({2A F + W A + W^2}) }$.  
Inserting these moments into Eq.
(\ref{eq:Rgrain})
we obtain

\begin{equation}
R_{grain} = { A F^2 \over {2A F + W A + W^2} }.  
\label{eq:RmomentP}
\end{equation}

\section{The limits of small and large grains}

Expressing the production rate 
$R_{grain}$
of Eq. 
(\ref{eq:RmomentP})
in terms of 
the size-independent hopping rate 
$a=AS$ (sec$^{-1}$)
and the incoming flux 
$f=F/S$ (monolayer sec$^{-1}$)
we obtain

\begin{equation}
R_{grain} = {{f S^2}\over
{\left({W\over f}\right)
\left[1+S \left( {W\over a}+{2f\over W}\right)\right]}}.
\label{eq:Rmoment}
\end{equation} 

\noindent
The ratio
$a/W$ is 
the number 
(up to a logarithmic correction)  
of sites that the atom visits before it desorbs
\citep{Montroll1965}.
The ratio
$W/f$ 
is roughly the average
number of vacant sites in the vicinity of an H atom.
Using these parameters we can identify two regimes 
of grain sizes.
For grains that are sufficiently large  
such that
$S > \min(a/W,W/f)$, 
the size dependence 
of the production rate is given by
$R_{\rm grain} \sim S \sim r^2$
and the production efficiency is independent of the grain size. 
This is the range in which rate equations apply.
In the limit of small grains, namely when
$S < \min(a/W,W/f)$, 
the production efficiency depends linearly on $S$,
and
$R_{\rm grain} \sim S^2 \sim r^4$.
This is the regime in which the grain size becomes the
smallest relevant length-scale in the system and
rate equations fail
\citep{Biham2002}. 

\begin{figure}
\epsfxsize=16cm
\epsffile{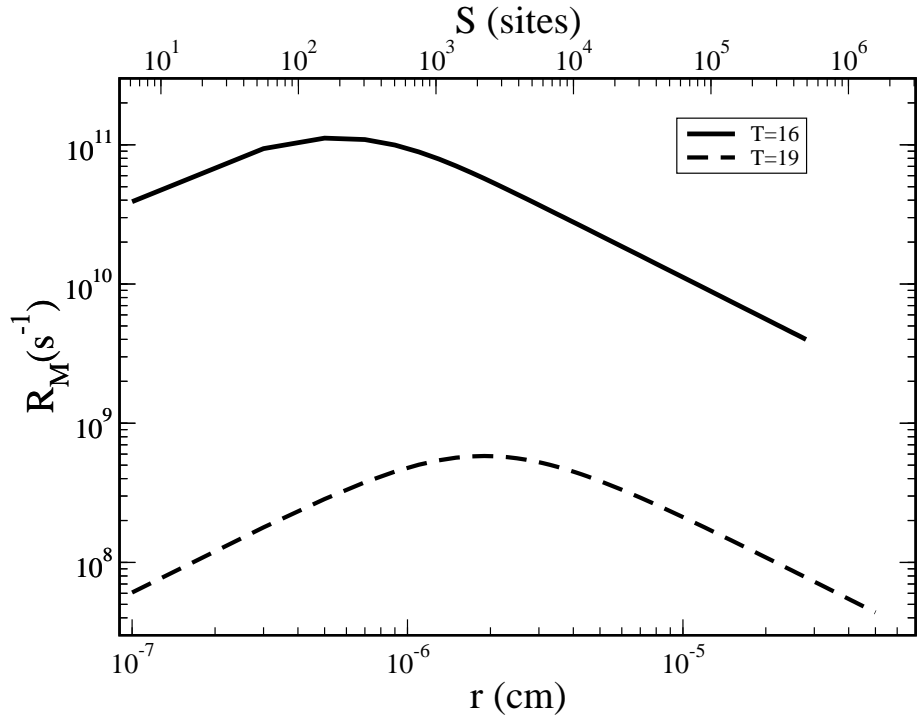}
\vspace{-4cm}
\caption{The total production rate $R_M$ (sec$^{-1}$) 
vs. grain radius $r$ for an ensemble of identical, spherical grains,
with total grain mass of
$M=1$ (gram). 
The density of the grain material is $\rho=2$ (gram cm$^{-3}$). 
The grain temperature is  
$T=16$ (solid line) 
and  
$T=19$ (dashed). 
The flux is 
$3.44 \times 10^{-9}$ (ML s$^{-1}$).
The optimal radius is 
$r_{\rm opt}=5.6 \times 10^{-7}$ (cm) 
for $T=16$K 
and  
$r_{\rm opt}=1.9 \times 10^{-6}$ (cm)
for 
$T=19$K.}
\label{fig:R}
\end{figure}

To illustrate the implications of these results,
consider a given mass $M$ of dust matter. 
Dividing the mass $M$ to spherical grains of radius $r$, 
the number of grains is given by
$n_{\rm grain} \sim r^{-3}$.
In an interstellar cloud, 
the $r$ dependence of the production rate of H$_2$ due to
such ensemble differs in the two limits discussed above.
For large grains
the total production rate is
$R_M = n_{\rm grain} R_{grain} \sim r^{-1}$. 
Reducing the grain size increases their total surface area, and 
as a result the production rate goes up.
In the limit of small grains the efficiency decreases as the
grain size is reduced and thus
$R_M \sim r$.
This indicates that there is an intermediate grain size
for which the production rate is maximal. 
In Fig. ~\ref{fig:R} we present the total production rate
$R_M$ (sec$^{-1}$) 
for a total dust mass of $M=1$ (gram)
which is divided into spherical grains of dradius $r$.
Clearly, there is an optimal grain size for which
the production rate is maximal. 
This optimal radius depends on
the grain composition and temperature and is given
by

\begin{equation}
r_{\rm opt}=\sqrt{S_{\rm opt}/4\pi s}
\label{eq:r_opt}
\end{equation}

\noindent
where 
$S_{\rm opt} = ({W/a}+2f/W)^{-1}$.
For typical interstellar conditions
this optimal grain radius is of the
order of tens of nm, 
with several thousand adsorption sites.

\section{The overall production rate of molecular hydrogen}

As discussed above, interstellar grains exhibit a broad size
distribution of the form 
$n_g(r) \sim r^{-\alpha}$. 
For 
$\alpha>2$
the surface area is dominated by the small grains,
while for 
$\alpha>3$
even the grain mass is dominated by the small grains.  
Therefore,
these small grains are expected to contribute significantly to H$_2$
formation. 
Taking this size distribution into account
we find that the production rate 
on grains with radii in the range $(r,r+dr)$ 
is given by
   
\begin{equation}
n_g(r)R_{\rm grain}(r)dr = \left\{
\begin{array}{ll}
c_1 r^{4-\alpha}dr: & \ {\rm small \ grains }  \\
c_2 r^{2-\alpha}dr: & \ {\rm large \ grains}
\end{array}
\right.
\label{eq:Rd}
\end{equation}

\noindent
where

\begin{equation}
c_1 = \frac{(4 \pi s f)^2 c}{W}
\label{eq:c1}
\end{equation} 

\noindent
and 

\begin{equation}
c_2 =  \frac{4 \pi s a f^2 c}{W^2 + 2af}
\label{eq:c2moment}
\end{equation}

\noindent
are obtained directly from 
Eq.
(\ref{eq:Rmoment})
in the limits of 
$S \ll S_{\rm opt}$
and
$S \gg S_{\rm opt}$,
respectively.
Eq. 
(\ref{eq:Rmoment})
is expected to be highly accurate in the limit of small grains.
For large grains its accuracy may be limited because its
derivation requires to impose a cutoff on the master equation.
For large grains the rate equations apply, and therefore
one can use

\begin{equation}
c_2={4\pi s W^2 c \over 16a}\left(-1+\sqrt{1+8{af \over W^2}}\right)^2,
\label{eq:c2rate}
\end{equation}

\noindent
which is obtained directly from the exact solution of the rate
equation
(\ref{eq:rate})
under steady state conditions
\citep{Biham1998}.

To evaluate the total production rate of molecular hydrogen 
per unit volume in an
interstellar cloud, one should take into account 
the size distribution of dust grains, 
and the dependence of the production efficiency 
on the grain size and temperature.
The total production rate is given by

\begin{equation}
R_{total}=\int_{r_{min}}^{r_{max}}n_g(r)R_{grain}(r)dr
\label{eq:integral}
\end{equation}

\noindent
where 
$n_g(r) dr$ is the density of 
grains with radii in the range  $(r,r+dr)$,
and 
$R_{grain}(r)$
can be taken either from the exact result
of Eq.
(\ref{eq:Rexact})
or from the
approximate result
of Eq.
(\ref{eq:Rmoment}).
Assuming a power law distribution of grain sizes, 
given by 
Eq. (\ref{eq:distribution}), 
and the result of
of Eq.
(\ref{eq:Rmoment})
one can divide the range of integration into
two parts,
namely the domain of small grains,
$r_{\rm min} < r <r_{\rm opt}$, 
and the domain of large grains,
$r_{\rm opt}<r<r_{\rm max}$.
Integrating each one of them separately, 
for $\alpha \neq 3$,
we obtain

\begin{equation}
R_{\rm total}={c_1 \over {5-\alpha}}\left(r_{\rm opt}^{5-\alpha}
-r_{\rm min}^{5-\alpha}\right)
+{c_2 \over {3-\alpha}}\left(r_{\rm max}^{3-\alpha}
-r_{\rm opt}^{3-\alpha}\right),
\label{eq:integration}
\end{equation}

\noindent
In the special case of $\alpha = 3$, the total production rate is

\begin{equation} 
R_{\rm total}={c_1 \over 2}\left(r_{\rm opt}^2-r_{\rm min}^2\right)+
c_2\ln \left({r_{\rm max}/{r_{\rm opt}}}\right).
\end{equation}

\begin{figure}
\epsfxsize=16cm
\epsffile{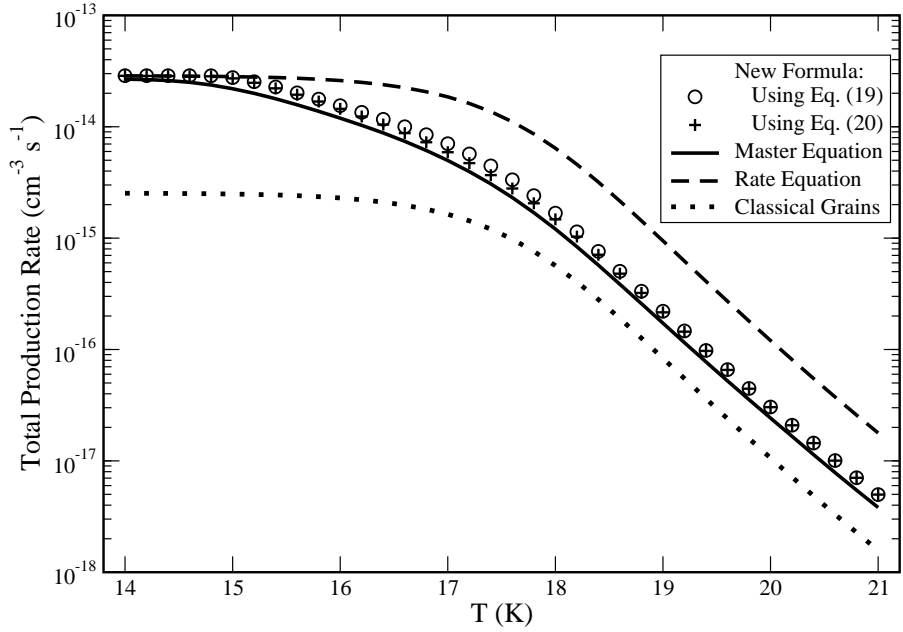}
\vspace{-4cm}
\caption{The total production rate 
$R_{\rm total}$ (cm$^{-3}$sec$^{-1}$)
of H$_2$ vs. grain temperature as calculated by the
formula 
(Eq.~\ref{eq:integration})
introduced in this paper,
where $c_2$ is given by Eq. 
(\ref{eq:c2moment})
($\circ$)
or by Eq.
(\ref{eq:c2rate})
($+$).
Both results are in very good agreement with the    
master equation (solid line),
although the latter one is slightly better.
The rate equation (dashed line) is found to 
over-estimate the production rate.
These results are compared to those obtained if the
same mass of grains is divided into identical, spherical grains 
of 'typical' radium ($r=0.17$ $\mu$ m) (dotted line).
It is found that for an astrophysically relevant range of high recombination
efficiency, the broad size distribution 
significantly enhances the production rate.}
\label{Fig:total}
\end{figure}

In 
Fig.~\ref{Fig:total} 
we present the total production rate obtained
from  
Eq.~(\ref{eq:integration})
for flux 
$f=6.88 \times 10^{-9}$ (ML sec$^{-1}$),
which corresponds to gas temperature of
90 K and hydrogen density 
$n_{\rm H} = 10$ (cm$^{-3}$).
The results of this formula are shown for the case in which
the parameter $c_2$ is taken from Eq. 
(\ref{eq:c2moment})
($\circ$)
and for the case in which it is taken from Eq.
(\ref{eq:c2rate})
($+$).
Both results
are in very good agreement with those obtained from the  
master equation (solid line).
The rate equation results (dashed line) 
are found to significantly over-estimate
the total production for most of the temperature range.
The results obtained  
(using the master equation)
under the assumption that the same total grain mass 
is divided into grains of identical size (0.17 $\mu m$)
are also shown (dotted line).
For the range of grain temperatures in which the H$_2$ production is 
efficient,
this assumption leads to an under-estimate of the total production rate
by about an order of magnitude.
This indicates that within this temperature range the broad size distribution
of the grains tends to enhance the production rate due to the increase in
grain-surface area.

\section{Discussion}

The formula introduced in this paper
[Eq. (\ref{eq:integration})]
enables us to calculate the formation rate of molecular hydrogen
in interstellar clouds for a broad range of conditions
without the need to perform computer simulations or numerical
integration. It applies both in the case of diffuse clouds in
which the H$_2$ formation takes place on bare silicate and carbon
grains and for molecular clouds in which the grains are
coated by ice mantles.
The formula applies directly for
grain size distributions that
can be approximated by a power-law of the
form of Eq.
(\ref{eq:distribution}).
For other distribution functions for which the integration cannot
be done analytically one can solve Eq.
(\ref{eq:integral})
by numerical integration.

The relaxation time for H$_2$ formation processes on dust grains
can be obtained from Eq.
(\ref{eq:rate}). It is given by
$\tau=(W^2+8AF)^{-1/2}$.
Under typical conditions,
this time is extremely short compared to the time scales
of interstellar clouds.
Therefore, the formula can be used even in the context of
evolving clouds. The time dependence can be taken into 
account by updating the input parameters to the formula
as the physical conditions in the cloud vary.

At very low grain temperatures below $T_0$, the mobility of H
atoms on the surface is suppressed and the rate of H$_2$ 
formation is sharply reduced. In this limit, H atoms keep
accumulating on the surface, the relaxation time $\tau$
increases and the behavior may no longer be well described
by steady state calculations. 
In this limit the coverage is relatively high and thus the
rate equations are valid.
However, the Langmuir rejection of atoms that hit the grain
in the vicinity of already adsorbed atoms should be
included. This is done by replacing the flux term $F$
by $F(1-N/S)$. 
The formula introduced in this paper,
Eq. (\ref{eq:integration}),
is valid only for grain temperatures higher than
$T_0$.
To extend its validity below 
$T_0$
one should incorporate the Langmuir rejection term into
it, by replacing all the appearances of $W$ in Eqs.
(\ref{eq:c1}) - (\ref{eq:c2rate})
by $W+f$.

In  
Eq. (\ref{eq:integration})
it is assumed that all the grains are at the
same temperature.
This assumption does not apply in photon dominated regions (PDR's),
where small grains exhibit large temperature fluctuations.
In general, the formation of molecular hydrogen in PDR's is
not well understood. 
The typical grain temperatures in these
regions is significantly higher than the temperature range in
which the formation of H$_2$ on
olivine, carbon and ice is found to be efficient.
Several approaches have been considered in attempt to explain 
this puzzle. 
One approach is based on the assumption that chemisorbed H atoms
play a role in the formation of H$_2$
\citep{Cazaux2002,Cazaux2004}.
Other explanations may ivolve porous grains that capture H atoms
more strongly within pores or with the temperature fluctuations
of small grains. 

\section{Summary}

We have presented a formula for the evaluation of the formation 
rate of molecular hydrogen in interstellar clouds, taking into
account the composition and temperature of the grains as well
as their size distribution.
The formula is found to be in excellent agreement with the
results of the master equation.
It enables us to obtain reliable results for the production rates
without the need to perform computer simulations or numerical
integrations.
It is found that for an astrophysically relevant range of grain
temperatures the broad size distribution of the grains provides
an enhancement of the H$_2$ formation rate compared to 
previous calculations.

\section{acknowledgments}
We thank E. Herbst, H.B. Perets, G. Vidali and V. Pirronello 
for helpful discussions.
This work was supported by the Israel Science Foundation
and the Adler Foundation for Space Research.

\clearpage

\bibliographystyle{unsrt}

\begin{thebibliography}{99}

\bibitem[\protect\citeauthoryear{Aronowitz \& Chang}
{1985}]{Aronowitz1985}
Aronowitz S., Chang S.B., 1985, ApJ.  293,  243  

\bibitem[\protect\citeauthoryear{Biham et al.}
{1998}]{Biham1998}
Biham O., Furman I., Katz, N., Pirronello, V.,  Vidali G.,  
1998, MNRAS., 296, 869 

\bibitem[\protect\citeauthoryear{Biham et al.}
{2001}]{Biham2001}
Biham O., Furman I., Pirronello, V.,  Vidali G.,  2001, ApJ. 553, 595 

\bibitem[\protect\citeauthoryear{Biham \& Lipshtat}{2002}]{Biham2002}
Biham O.,  Lipshtat A., 2002, Phys. Rev. E, 66, 056103

\bibitem[\protect\citeauthoryear{Brown \& Charnley}{1990}]{Brown1990b}
Brown P.D.,  Charnley S.B., 1990, MNRAS, 244, 432

\bibitem[\protect\citeauthoryear{Caselli et al.}{1993}]{Caselli1993}
Caselli, P.,  Hasegawa T.I.,  Herbst E., 1993, ApJ, 408, 548

\bibitem[\protect\citeauthoryear{Caselli et al.}{1998}]{Caselli1998}
{Caselli, P.,  Hasegawa, T.I., \&  Herbst, E.} 1998, ApJ, 495, 309

\bibitem[\protect\citeauthoryear{Cazaux \& Tielens}{2002}]{Cazaux2002}
Cazaux S., Tielens A. G. G. M., 2002, ApJ, 575, L29

\bibitem[\protect\citeauthoryear{Cazaux \& Tielens}{2004}]{Cazaux2004}
Cazaux S., Tielens A. G. G. M., 2004, ApJ, 604, 222

\bibitem[\protect\citeauthoryear{Charnley et al.}{1997}]{Charnley1997}
Charnley S.B., Tielens A.G.G.M.,  Rodgers S.D., 1997, ApJ, 482, L203

\bibitem[\protect\citeauthoryear{Draine \& Lee}{1984}]{Draine1984}
Draine B.T.,  Lee H.M., 1984, ApJ, 285,  89

\bibitem[\protect\citeauthoryear{Duley \& Williams}{1986}]{Duley1986}
Duley W.W.,  Williams D.A., 1986, MNRAS 223, 177

\bibitem[\protect\citeauthoryear{Gould \& Salpeter}{1963}]{Gould1963}
Gould R.J., Salpeter E.E., 1963, ApJ, 138, 393

\bibitem[\protect\citeauthoryear{Green et al.}{2001}]{Green2001}
Green N.J.B., Toniazzo T., Pilling M.J. Ruffle, D.P. Bell, N., 
Hartquist T.W., 2001, A\&A 375, 1111 

\bibitem[\protect\citeauthoryear{Hartquist and  Williams}{1995}]{Hartquist1995}
Hartquist T.W.,  Williams D.A., 1995, 
{\em The chemically controlled cosmos}, ({Cambridge University Press} {Cambridge, UK})

\bibitem[\protect\citeauthoryear{Hasegawa et al.}{1992}]{Hasegawa1992}
Hasegawa T.I., Herbst E., Leung C.M., 1992, ApJS,  82,  167

\bibitem[\protect\citeauthoryear{d'Hendecourt et al.}{1985}]{Hendecourt1985}
d'Hendecourt L.B., Allamandola L.J., Greenberg J.M., 1985, A\&A, 152, 
130

\bibitem[\protect\citeauthoryear{Hollenbach \& Salpeter}{1970}]{Hollenbach1970}
Hollenbach D., Salpeter E.E., 1970, J. Chem. Phys. 53, 79 

\bibitem[\protect\citeauthoryear{Hollenbach \& Salpeter}{1971}]{Hollenbach1971a}
Hollenbach D.,  Salpeter E.E., 1971,  ApJ 163, 155

\bibitem[\protect\citeauthoryear{Hollenbach et al.}{1971}]{Hollenbach1971b}
Hollenbach D., Werner M.W., Salpeter E.E., 1971, ApJ 163,  165

\bibitem[\protect\citeauthoryear{Katz et al.}{1999}]{Katz1999}
Katz N., Furman I., Biham O., Pirronello V., Vidali G., 1999,  ApJ, 
522, 305

\bibitem[\protect\citeauthoryear{Lipshtat \& Biham}{2003}]{Lipshtat2003}
Lipshtat A. and Biham O., 2003, A\&A, 400, 585

\bibitem[\protect\citeauthoryear{Mathis et al.}{1977}]{Mathis1977}
Mathis J.S., Rumpl W., Nordsieck K.H., 1977, ApJ, 217, 425

\bibitem[\protect\citeauthoryear{Montroll \& Weiss}{1965}]{Montroll1965}
Montroll E.W.,  Weiss G.H., 1965, J. Math. Phys. 6, 167 

\bibitem[\protect\citeauthoryear{Pickles \& Williams}{1977}]{Pickles1977}
Pickles J.B., Williams D.A., 1977, Ap\&SS, 52,  433

\bibitem[\protect\citeauthoryear{Pirronello \& Avera}{1988}]{Pirronello1988}
Pirronello V.,  Averna D.. 1988, A\&A, 201, 196

\bibitem[\protect\citeauthoryear{Pirronello et al.}{1997a}]{Pirronello1997a}
Pirronello V., Liu C.,  Shen L.,  Vidali G., 1997a, ApJ 475, L69

\bibitem[\protect\citeauthoryear{Pirronello et al.}{1997b}]{Pirronello1997b}
Pirronello V., Biham O., Liu C., Shen L., Vidali G., 1997b, ApJ  483, 
L131

\bibitem[\protect\citeauthoryear{Pirronello et al.}{1999}]{Pirronello1999}
Pirronello V., Liu C., Roser J.E., Vidali G., 1999, A\&A 344, 681

\bibitem[\protect\citeauthoryear{Roser et al.}{2002}]{Roser2002}
Roser, J.E., Manic\'o, G., Pirronello, V., Vidali, G., 2002, ApJ 581, 276 

\bibitem[\protect\citeauthoryear{Sandford \& Allamandola}{1993}]{Sandford1993} 
Sandford S.A.,  Allamandola L.J., 1993, ApJ, 409, L65  

\bibitem[\protect\citeauthoryear{Shalabiea et al.}{1998}]{Shalabiea1998}
Shalabiea O.M., Caselli P., Herbst E., 1998, ApJ, 502, 652

\bibitem[\protect\citeauthoryear{Smoluchowski}{1981}]{Smoluchowski1981}
Smoluchowski R., 1981, Ap\&SS, 75, 353

\bibitem[\protect\citeauthoryear{Smoluchowski}{1983}]{Smoluchowski1983}
Smoluchowski R., 1983, J. Chem. Phys., 87, 4229

\bibitem[\protect\citeauthoryear{Stantcheva et al.}{2001}]{Stantcheva2001}
Stantcheva T., Caselli P.,  Herbst E., 2001, A\&A, 375, 673

\bibitem[\protect\citeauthoryear{Weingartner}{2001}]{Weingartner2001}
Weingartner J.C., Draine B.T., 2001, ApJ, 548, 296 

\bibitem[\protect\citeauthoryear{Willacy \& Williams}{1993}]{Willacy1993}
Willacy K.,  Williams D.A., 1993, MNRAS, 260, 635

\bibitem[\protect\citeauthoryear{Williams}{1968}]{Williams1968}
Williams D.A., 1968, ApJ, 151, 935

\end{thebibliography}

\clearpage

\end{document}